\documentclass[preprint,showpacs,amsmath,amssymb,aps,prd,nofootinbib]{revtex4}
\usepackage{epsfig,graphicx,color,appendix,fge}
\begin{document}
%\begin{flushright}
%IBS-CTPU-18-07
%\end{flushright}
\def\CP{{\it CP}~}
\def\cp{{\it CP}}
\title{\mbox{}\\[10pt]
Compact model\footnote{Here `compact' model means a model that provides only requisite parameters it is easy to disprove.} for Quarks and Leptons \\via flavored-Axions}

\author{Y. H. Ahn}
\affiliation{Key Laboratory of Particle Astrophysics, Institute of High Energy Physics,
Chinese Academy of Sciences, Beijing, 100049, China}
\email{axionahn@naver.com}

%\thanks{}


%\date{\today}

%%%%%%%%%%%%%%%%%%%%%%%%%%%%%%%%%%%%%%%%%%%%%%%%%%%%%%%%%%%%%%%%%%%%%%%%%%%%%%%%%%%%%%%%%%
\begin{abstract}
We show how the scales responsible for Peccei-Quinn (PQ), seesaw, and Froggatt and Nielsen (FN) mechanisms can be fixed, by constructing a compact model for resolving rather recent, but fast-growing issues in astro-particle physics, including quark and leptonic mixings and CP violations, high-energy neutrinos, QCD axion, and axion cooling of stars. The model is motivated by the flavored PQ symmetry for unifying the flavor physics and string theory.
The QCD axion decay constant congruent to the seesaw scale, through its connection to the astro-particle constraints of both the stellar evolution induced by the flavored-axion bremsstrahlung off electrons $e+Ze\rightarrow Ze+e+A_i$ and the rare flavor-changing decay process induced by the flavored-axion $K^+\rightarrow\pi^++A_i$, is shown to be fixed at $F_A=3.56^{+0.84}_{-0.84}\times10^{10}$ GeV (consequently, the QCD axion mass $m_a=1.54^{+0.48}_{-0.29}\times10^{-4}$ eV, Compton wavelength of its oscillation $\lambda_a=8.04^{+1.90}_{-1.90}\,{\rm mm}$, and axion to neutron coupling $g_{Ann}=2.14^{+0.66}_{-0.41}\times10^{-12}$, etc.). Subsequently, the scale associated to FN mechanism is dynamically fixed, $\Lambda=2.04^{\,+0.48}_{\,-0.48}\times10^{11}\,{\rm GeV}$, through its connection to the standard model fermion masses and mixings, and such fundamental scale might give a hint where some string moduli are stabilized in type-IIB string vacua. In the near future, the NA62 experiment expected to reach the sensitivity of ${\rm Br}(K^+\rightarrow\pi^++A_i)<1.0\times10^{-12}$ will probe the flavored-axions or exclude the model, if the astrophysical constraint of star cooling is really responsible for the flavored-axion.

\end{abstract}

\maketitle %
%%%%%%%%%%%%%%%%%%%%%%%%%%%%%%%%%%%%%%%%%%%%%%%%%%%%%%%%%%%%%%%%%%%%%%%%%%%%%%%
\section{Introduction}
%For all the success of the Standard Model (SM), it is on the verge of being surpassed. 
Until now, symmetries have played an important role in physics in general and in quantum field theory in particular.
The SM as a low-energy effective theory has been very predictive and well tested, due to the symmetries satisfied by the theory - Lorentz invariance plus the $SU(3)_C\times SU(2)_L\times U(1)_Y$ gauge symmetry in addition to the discrete space-time symmetries like P and CP. However, it leaves many open questions for theoretical and cosmological issues that have not been solved yet ({\it e.g.}, \cite{Ahn:2016hbn, Ahn:2017dpf}).
The SM therefore cannot be the final answer. It is widely believed that the SM should be extended to a more fundamental underlying theory.
Neutrino mass and mixing is the first new physics beyond SM and adds impetus to solving several open questions in astro-particle physics and cosmology. Seesaw mechanism\,\cite{Minkowski:1977sc} has been the most promising one responsible for the neutrino mass. Moreover, a solution to the strong CP problem of QCD through Peccei-Quinn (PQ) mechanism\,\cite{Peccei-Quinn}\,\footnote{See, most recent its related simple toy models ((non-)supersymmetric versions) in Ref\,.\,\cite{Ahn:2015pia}, see also Ref.\,\cite{Cheng:1995fd}.} may hint a new extension of gauge theory\,\cite{Ahn:2016typ, Ahn:2016hbn}. If the QCD axion as a solution to the strong CP problem exists, it can easily fit into a string theoretic framework and appears cosmologically as a form of cold dark matter\,\footnote{On this issue we will consider flavored-axion\,\cite{Ahn:2014gva} as a cold dark matter in the next project. The scale in Eq.\,(\ref{k_bound}) we found is available for explaining dark matter, and finding it in experiments can change the fundamental understanding of the universe.}. Flavor puzzle for the SM charged-fermion mass hierarchies could be solved by implementing Froggatt and Nielsen (FN) mechanism\,\cite{Froggatt:1978nt}. If those mechanisms are realized in nature at low energies, finding the scales responsible for the seesaw, PQ, and  FN mechanisms could be one of big challenges as a theoretical guideline to the fundamental issues of particle physics and cosmology. 

Many of the outstanding mysteries of astrophysics may be hidden from our sight at all wavelengths of the electromagnetic spectrum because of absorption by matter and radiation between us and the source. So, data from a variety of observational windows, especially, through direct observations with neutrinos and axions, may be crucial. Hence, axions and neutrinos in astro-particle physics and cosmology could be powerful sources for a new extension of SM particle physics\,\cite{Ahn:2016hbn, Ahn:2017dpf, Ahn:2016typ}, in that they stand out as their convincing physics and the variety of experimental probes.
Fortunately, most recent analyses on the knowledges of neutrino (low-energy neutrino oscillations\,\cite{Esteban:2016qun} and high-energy neutrino\,\cite{Aartsen:2013jdh}) and axion (QCD axion\,\cite{Asztalos:2009yp, Asztalos:2003px} and axion-like-particle(ALP)\,\cite{Giannotti:2017hny, Ringwald:2015lqa}) enter into a new phase of model construction for quarks and leptons\,\cite{Ahn:2014gva}. In light of finding the fundamental scales, interestingly enough, there are two astro-particle constraints coming from the star cooling induced by the flavored-axion bremsstrahlung off electrons $e+Ze\rightarrow Ze+e+A_i$\,\cite{Giannotti:2017hny} and the rare flavor-changing decay process induced by the flavored-axion $K^+\rightarrow\pi^++A_i$\,\cite{Adler:2008zza}, respectively,
\begin{eqnarray}
 6.7\times10^{-29}\lesssim\alpha_{Aee}\lesssim5.6\times10^{-27}\quad\text{at}~3\sigma\,,\qquad {\rm Br}(K^+\rightarrow\pi^+A_i)<7.3\times10^{-11}\,,
 \label{AP_cons}
\end{eqnarray}
where $\alpha_{Aee}$ is the fine-structure of axion to electron.
 Since astro-particle physics observations have increasingly placed tight constraints on parameters for flavored-axions, it is in time for a compact model for quarks and leptons to mount an interesting challenge on fixing the fundamental scales such as the scales of seesaw, PQ, and FN mechanisms.
The purpose of the present paper is to construct a flavored-PQ model\,\cite{Ahn:2014gva} along the lines of the challenge, which naturally extends to a compact symmetry $G_F=$ {\it anomalous $U(1)$} plus {\it non-Abelian (finite) symmetries} for new physics beyond SM. Remark that\,\cite{Ahn:2018nfb} in modeling the $U(1)$ mixed-gravitational anomaly cancellation\,\cite{Green:1984sg} is of central importance in constraining the fermion contents of a new chiral gauge theory and the flavor structure of $G_F$ is strongly correlated with physical observables. Here the flavored-PQ $U(1)$ symmetry together with the non-Abelian finite symmetry is well flavor-structured in a unique way that domain-wall number $N_{\rm DW}=1$ with the $U(1)_X\times[gravity]^2$ anomaly-free condition demands additional Majorana fermion and the flavor puzzles of SM are well delineated by new expansion parameters expressed in terms of $U(1)_X$ charges and $U(1)_X$-$[SU(3)_C]^2$ anomaly coefficients, providing interesting physical implications on neutrino, QCD axion, and flavored-axion\,\footnote{Recently, studies on flavored-axion are gradually becoming amplified\,\cite{Ahn:2014gva, Ahn:2016hbn, Ahn:2018nfb, flavoraxion}.}.

The rest of this paper is organized as follows. In Sec.\,II we construct a compact model based on $SL_2(F_3)\times U(1)_X$ in a supersymmetric framework. 
Subsequently, we show that the model works well with the SM fermion mass spectra and their peculiar flavor mixing patterns.  
In Sec.\,III we show that the QCD decay constant (congruent to the seesaw scale) is well fixed through constraints coming from astro-particle physics, and in turn the FN scale is dynamically determined via its connection to the SM fermion masses and mixings. And we show several properties of the flavored-axions.
What we have done is summarized in Sec.\,V. 

%%%%%%%%%%%%%%%%%%%%%%%%%%%%%%%%%%%%%%%%%%%%%%%%%%%%%%%%%%%%%%%%%%%%%%%%%%%%%%%%%%%%%%%%%%
\section{flavored $SL_{2}(F_3)\times U(1)_{X}$ symmetry}

Similar to Ref.\,\cite{Ahn:2018nfb}, assume that we have a SM gauge theory based on the $G_{\rm SM}=SU(3)_C\times SU(2)_L\times U(1)_Y$ gauge group, and that the theory has in addition a $G_F\equiv SL_2(F_3)\times U(1)_X$ for a {\it compact} description of new physics beyond SM. Here the symmetry group of the double tetrahedron $SL_2(F_3)$\,\cite{slf3, aranda, Feruglio:2007uu} could be realized in field theories on orbifolds; it is a subgroup of a gauge symmetry that can be protected from quantum-gravitational effects.
 And the $U(1)_X$ as flavored-PQ symmetry is composed of two anomalous symmetries $U(1)_{X_1}\times U(1)_{X_2}$ generated by the charges $X_1\equiv-2p$ and $X_2\equiv-q$.  Here the global $U(1)$ symmetry\,\footnote{It is likely that an exact continuous global symmetry is violated by quantum gravitational effects\,\cite{Krauss:1988zc}.} including $U(1)_R$ is remnants of the broken $U(1)$ gauge symmetries which can connect string theory with flavor physics\,\cite{Ahn:2016typ, Ahn:2016hbn}. Hence, the spontaneous breaking of $U(1)_X$ realizes the existence of the Nambu-Goldstone (NG) mode (called axion) and provides an elegant solution to the strong CP problem.

%%%%%%%%%%%%%%%%%%%%%%%%%%%%%%%%%%%%%%%%%%%%%%%%%%%%%%%%%%%%%%%%%%%%%%%%%%%%%%%
\subsection{Vacuum configuration}
\label{vacCon}
\noindent We briefly review the fields contents responsible for the vacuum configuration since the scalar potential of the model is the same as in Ref.\,\cite{Ahn:2018nfb}.
Apart from the usual two Higgs doublets $H_{u,d}$ responsible for electroweak symmetry breaking, which transform as $(\mathbf{1},0)$ under $SL_2(F_3)\times U(1)_X$ symmetry, the scalar sector is extended via two types of new scalar multiplets that are $G_{\rm SM}$-singlets: flavon fields $\Phi_{T},\Phi_{S},\Theta,\tilde{\Theta}, \eta, \Psi, \tilde{\Psi}$ responsible for the spontaneous breaking of the flavor symmetry, and driving fields $\Phi^{T}_{0},\Phi^S_{0},\eta_0, \Theta_{0},\Psi_{0}$ that are to break the flavor group along required VEV directions and to allow the flavons to get VEVs, which couple only to the flavons. The electroweak Higgs fields $H_{u,d}$ are enforced to be neutral under $U(1)_X$ not to have an axionic domain-wall problem.

Under $SL_2(F_3)\times U(1)_X$ the flavon fields $\{\Phi_{T},\Phi_{S}\}$ transform as $(\mathbf{3}, 0)$ and $(\mathbf{3}, X_1)$, $\eta$ as $(\mathbf{2}', 0)$, and $\{\Theta,\tilde{\Theta},\Psi,\tilde{\Psi}\}$ as $(\mathbf{1}, X_1)$, $(\mathbf{1}, X_1)$, $(\mathbf{1}, X_2)$, and $(\mathbf{1}, -X_2)$, respectively; the driving fields $\{\Phi_{0}^{T},\Phi_{0}^{S}\}$ transform as $(\mathbf{3}, 0)$ and $(\mathbf{3}, -2X_1)$, $\eta_0$ as $(\mathbf{2}'', 0)$, and $\{\Theta_{0}, \Psi_{0}\}$ as $(\mathbf{1}, -2X_1)$ and $(\mathbf{1}, 0)$, respectively.
For vacuum stability and a desired vacuum alignment solution, the flavon fields $\{\Phi_T, \eta\}$ are enforced to be neutral under $U(1)_X$. 
In addition, the superpotential $W$ in the theory is uniquely determined by the $U(1)_R$ symmetry, containing the usual $R$-parity as a subgroup: $\{matter\,fields\rightarrow e^{i\xi/2}\,matter\,fields\}$ and $\{driving\,fields\rightarrow e^{i\xi}\,driving\,fields\}$, with $W\rightarrow e^{i\xi}W$, whereas flavon and Higgs fields remain invariant under an $U(1)_R$ symmetry. %As a consequence of the $R$ symmetry, the other superpotential term $\kappa_{\alpha}L_{\alpha}H_{u}$ and the terms violating the lepton and baryon number symmetries are not allowed. In addition, dimension 6 supersymmetric operators like $Q_{i}Q_{j}Q_{k}L_{l}$ ($i,j,k$ must not all be the same) are not allowed either, and stabilizing proton.
As Ref.\,\cite{Ahn:2018nfb} the global minima of the potential are given at leading order by
{\small\begin{eqnarray}
 &&\langle\Phi_T\rangle=\frac{v_{T}}{\sqrt{2}}(1,0,0)\,,\qquad\qquad\langle\Phi_S\rangle=\frac{v_{S}}{\sqrt{2}}(1,1,1)\,,\qquad\qquad\langle\eta\rangle=\frac{v_{\eta}}{\sqrt{2}}(1,0)\,,\nonumber\\
 &&\qquad\quad\langle\Psi\rangle=\langle\tilde{\Psi}\rangle=\frac{v_\Psi}{\sqrt{2}}\,, \qquad\qquad\langle\Theta\rangle=\frac{v_\Theta}{\sqrt{2}}\,,\qquad\qquad \langle\tilde{\Theta}\rangle=0\,,
  \label{vev}
\end{eqnarray}}
where $v_\Psi=v_{\tilde{\Psi}}$ and $\kappa=v_S/v_\Theta$  in SUSY limit. 
And the complex scalar fields are decomposed as follows\,\cite{Ahn:2018nfb}
 \begin{eqnarray}
  &&\Phi_{Si}=\frac{e^{i\frac{\phi_{S}}{v_{S}}}}{\sqrt{2}}\left(v_{S}+h_{S}\right)\,,\qquad\qquad\quad~\Theta=\frac{e^{i\frac{\phi_{\theta}}{v_{\Theta}}}}{\sqrt{2}}\left(v_{\Theta}+h_{\Theta}\right)\,,\nonumber\\
&&\Psi=\frac{v_{\Psi}}{\sqrt{2}}e^{i\frac{\phi_{\Psi}}{v_{g}}}\left(1+\frac{h_{\Psi}}{v_{g}}\right)\,,\qquad\qquad\,\tilde{\Psi}=\frac{v_{\tilde{\Psi}}}{\sqrt{2}}e^{-i\frac{\phi_{\Psi}}{v_{g}}}\left(1+\frac{h_{\tilde{\Psi}}}{v_{g}}\right)\,,
  \label{NGboson}
 \end{eqnarray}
in which $\Phi_{S1}=\Phi_{S2}=\Phi_{S3}\equiv\Phi_{Si}$ and $h_{\Psi}=h_{\tilde{\Psi}}$ in the SUSY limit, and $v_{g}=\sqrt{v^2_{\Psi}+v^2_{\tilde{\Psi}}}$. And the NG modes $A_1$ and $A_2$ are expressed as
 \begin{eqnarray}
  A_1=\frac{v_{S}\,\phi_{S}+v_{\Theta}\,\phi_{\theta}}{\sqrt{v^{2}_{S}+v^{2}_{\Theta}}}\,,\qquad A_{2}=\phi_{\Psi}
 \end{eqnarray}
with the angular fields $\phi_{S}$, $\phi_{\theta}$ and $\phi_{\Psi}$.

%%%%%%%%%%%%%%%%%%%%%%%%%%%%%%%%%%%%%%%%%%%%%%%%%%%%%%%%%%%%%%%%%%%%%%%%%%%%%
\subsection{Quarks, Leptons, and flavored-Axions}
\label{qla}
\noindent Under $SL_2(F_3)\times U(1)_{X}$ with $U(1)_R=+1$, the SM quark matter fields are sewed by the five (among seven) in-equivalent representations ${\bf 1}$, ${\bf 1}'$, ${\bf 1}''$, ${\bf 2}'$ and ${\bf 3}$ of $SL_2(F_3)$, and assigned as in Table\,\ref{reps_q} and \ref{reps_l}. Because of the chiral structure of weak interactions, bare fermion masses are not allowed in the SM. Fermion masses arise through Yukawa interactions\,\cite{Ahn:2014zja}. 
%\begin{center}
\begin{table}[h]
\caption{\label{reps_q} Representations of the quark fields under $SL_2(F_3)\times U(1)_{X}$ with $U(1)_R=+1$.}
\begin{ruledtabular}
\begin{tabular}{cccc}
Field &$Q_{1},~Q_{2},~Q_{3}$&${\cal D}^c, ~b^c$&${\cal U}^c, ~t^c$\\
\hline
$SL_2(F_3)$&$\mathbf{1}$, $\mathbf{1}'$, $\mathbf{1^{\prime\prime}}$&$\mathbf{2}'$, $\mathbf{1}'$&$\mathbf{2}'$, $\mathbf{1^{\prime}}$\\
$U(1)_{X}$&$10p-4q,~8p-2q,~0$ &$3q-8p$, ~$3q$&$-8p,~0$\\
%$U(1)_R$&$1$ &~$1$~&~$1$\\
%$SU(2)\times U(1)_Y$&$2_{-1}$&$2_\frac{1}{3}$&$1_{\frac{4}{3}}$&$1_{-\frac{2}{3}}$&$1_{-2}$&$1_{0}$&$2_{1}$&$2_{-1}$\\
\end{tabular}
\end{ruledtabular}
\end{table}
%\end{center} 
Then the Yukawa superpotential for quark sector invariant under $G_{\rm SM}\times G_F\times U(1)_R$ is sewed as
%\begin{widetext}
{\begin{eqnarray}
 W^u_q &=&
  \hat{y}_{t}\,t^cQ_{3}H_u
  +y_{c}\,(\eta {\cal U}^c)_{{\bf 1}''}Q_2\frac{H_u}{\Lambda}+\tilde{y}_{c}\,[(\eta {\cal U}^c)_{{\bf 3}}\Phi_T]_{{\bf 1}''}Q_2\frac{H_u}{\Lambda^2}\nonumber\\
  &+&y_{u}\,[(\eta {\cal U}^c)_{\bf 3}\Phi_T]_{\bf 1}Q_1\frac{H_u}{\Lambda^2}+\tilde{y}_{u}\,[(\eta {\cal U}^c)_{\bf 3}\eta\eta]_{\bf 1}Q_1\frac{H_u}{\Lambda^3}\,,\label{lagrangian_qu}\\
%   \end{eqnarray}}
% {\begin{eqnarray}
 W^d_q &=& y_{b}\,b^cQ_{3}H_d
  +y_{s}\,(\eta {\cal D}^c)_{{\bf 1}''}Q_2\frac{H_d}{\Lambda}+Y_{s}\,b^cQ_2(\Phi_S\Phi_S)_{{\bf 1}'}\frac{H_d}{\Lambda^2}+y_{d}\,[(\eta {\cal D}^c)_{\bf 3}\Phi_S]_{{\bf 1}}Q_{1}\frac{H_d}{\Lambda^2}\nonumber\\
  &+& Y_{d}\,b^cQ_1(\Phi_S\Phi_S)_{{\bf 1}''}\frac{H_d}{\Lambda^2}
  + \tilde{y}_{d}\,[(\eta {\cal D}^c)_{\bf 3}\Phi_T]_{{\bf 1}}Q_{1}\frac{H_d}{\Lambda^2}\,.
 \label{lagrangian_qd}
 \end{eqnarray}}
%\end{widetext}
According to the assignment of the $U(1)_{X}$ quantum numbers to the matter fields content as in Table\,\ref{reps_q}, the Yukawa couplings of quark fermions are visualized as a function of the SM gauge singlet flavon fields $\Psi(\tilde{\Psi})$ and /or $\Theta (\Phi_S)$, except for the top Yukawa coupling:
\begin{eqnarray}
y_c &=&\hat{y}_c\Big(\frac{\tilde{\Psi}}{\Lambda}\Big)^2\,,\qquad\tilde{y}_c =\hat{\tilde{y}}_c\Big(\frac{\tilde{\Psi}}{\Lambda}\Big)^2\,,\qquad y_u=\hat{y}_u\Big(\frac{\tilde{\Psi}}{\Lambda}\Big)^4\frac{\Theta}{\Lambda}\,,\qquad \tilde{y}_u=\hat{\tilde{y}}_u\Big(\frac{\tilde{\Psi}}{\Lambda}\Big)^4\frac{\Theta}{\Lambda}\,,\nonumber\\
y_b&=&\hat{y}_b
\Big(\frac{\Psi}{\Lambda}\Big)^3\,,\qquad y_s=\hat{y}_s\Big(\frac{\Psi}{\Lambda}\Big)\,,\,~\qquad y_d=\hat{y}_d\Big(\frac{\tilde{\Psi}}{\Lambda}\Big)\,,~\,\quad\qquad \tilde{y}_d=\hat{\tilde{y}}_d\Big(\frac{\tilde{\Psi}}{\Lambda}\Big)\frac{\Theta}{\Lambda}\,,\nonumber\\
Y_s&=&\hat{Y}_{s1}\Big(\frac{\Theta}{\Lambda}\Big)^2\frac{\Psi}{\Lambda}+\hat{Y}_{s2}\Big(\frac{\Phi_S}{\Lambda}\Big)^2\frac{\Psi}{\Lambda}\,,\qquad\qquad Y_d=\hat{Y}_{d1}\Big(\frac{\Theta}{\Lambda}\Big)^{3}\frac{\tilde{\Psi}}{\Lambda}+\hat{Y}_{d2}\Big(\frac{\Phi_S}{\Lambda}\Big)^{2}\frac{\Theta}{\Lambda}\frac{\tilde{\Psi}}{\Lambda}\,.
\end{eqnarray}
Recalling that the hat Yukawa coupling denotes order of unity.
The up-type quark superpotential in Eq.\,(\ref{lagrangian_qu}) does not contribute to the Cabbibo-Kobayashi-Maskawa (CKM) matrix due to the diagonal form of mass matrix, while the down-type quark superpotential in Eq.\,(\ref{lagrangian_qd}) does contribute the CKM matrix.

As discussed in Refs.\,\cite{Ahn:2018nfb, Ahn:2016hbn, Ahn:2014gva}, with the condition of $U(1)_X$-$[gravity]^2$ anomaly cancellation new additional Majorana fermions $S^{c}_{e,\,\mu,\,\tau}$ besides the heavy Majorana neutrinos can be introduced in the lepton sector. Hence, such new additional Majorana neutrinos can play a role of the active neutrinos as pseudo-Dirac neutrinos.
Under $SL_2(F_3)\times U(1)_{X}$ with $U(1)_R=+1$, the quantum numbers of the lepton fields are summarized as in Table\,\ref{reps_l}.
%\begin{center}
\begin{table}[h]
\caption{\label{reps_l} Representations of the lepton fields under $SL_2(F_3)\times U(1)_{X}$ with $U(1)_R=+1$. And here $r\equiv {\cal Q}_{y_{\nu}}+p$ is defined.}
\begin{ruledtabular}
\begin{tabular}{ccccc}
Field &$L$&$e^c,\mu^c,\tau^c$&$N^{c}$&$S_e^c,S_\mu^c,S_\tau^c$\\
\hline
$SL_2(F_3)$&$\mathbf{3}$&$\mathbf{1}$, $\mathbf{1^{\prime\prime}}$, $\mathbf{1^\prime}$&$\mathbf{3}$&$\mathbf{1}$, $\mathbf{1^{\prime\prime}}$, $\mathbf{1^{\prime}}$\\
$U(1)_{X}$& $ -r $ & $r-{\cal Q}_{y_e}, r-{\cal Q}_{y_\mu}, r-{\cal Q}_{y_\tau}$& $p$&$r-{\cal Q}_{y^s_1}$, $r-{\cal Q}_{y^s_2}$, $r-{\cal Q}_{y^s_3}$\\
%$U(1)_R$& $ 1 $ & $1$~~~& $1$& $1$\\
%$SU(2)\times U(1)_Y$&$2_{-1}$&$2_\frac{1}{3}$&$1_{\frac{4}{3}}$&$1_{-\frac{2}{3}}$&$1_{-2}$&$1_{0}$&$2_{1}$&$2_{-1}$\\
\end{tabular}
\end{ruledtabular}
\end{table}
%\end{center} 
The lepton Yukawa superpotential, similar to the quark sector, invariant under $G_{\rm SM}\times G_F\times U(1)_R$ reads at leading order
%\begin{widetext}
{\begin{eqnarray}
 W_{\ell} &=&
y_\tau\,\tau^c (L\Phi_T)_{{\bf 1}''} \frac{H_d}{\Lambda}+y_\mu\,\mu^c (L\Phi_T)_{{\bf 1}'} \frac{H_d}{\Lambda}+y_e\,e^c (L\Phi_T)_{{\bf 1}} \frac{H_d}{\Lambda}\,,\label{lagrangian_chL}\\
%   \end{eqnarray}}
%  {{\begin{eqnarray}
 W_{\nu} &=&y^s_3\,S^c_\tau(L\Phi_T)_{{\bf 1}''} \frac{H_u}{\Lambda}+y^s_2\, S^c_\mu(L\Phi_T)_{{\bf 1}'} \frac{H_u}{\Lambda}+y^s_1\,S^c_e (L\Phi_T)_{{\bf 1}} \frac{H_u}{\Lambda}\nonumber\\
 &+& y_{\nu}(LN^c)_{{\bf 1}}H_{u}+\frac{1}{2}(\hat{y}_\Theta\Theta+\hat{y}_{\tilde{\Theta}}\tilde{\Theta})(N^{c}N^{c})_{{\bf 1}}+\frac{\hat{y}_R}{2}(N^{c}N^{c})_{{\bf 3}} \Phi_S\nonumber\\  
 &+&\frac{1}{2}\{y^{ss}_1\,S^c_eS^c_e+y^{ss}_2\,S^c_\mu S^c_\tau+y^{ss}_2\,S^c_\tau S^c_\mu\}\Theta\,.
 \label{lagrangian2}
 \end{eqnarray}}
%\end{widetext}
Below the cutoff scale $\Lambda$, the mass term of the Majorana neutrinos $N^c$ comprises an exact tri-bimaximal mixing (TBM) pattern\,\cite{Harrison:2002er, Altarelli:2005yp}.
%Imposing the $U(1)_{X}$ symmetry explains the absence of the Yukawa terms $LN^{c}\Phi_{S}$ and $N^{c}N^{c}\Phi_{T}$ as well as does not allow the interchange between $\Phi_{T}$ and $\Phi_{S}$, both of which transform differently under $U(1)_{X}$, so that the exact TBM is obtained at leading order. 
With the desired VEV alignment in Eq.\,(\ref{vev}) it is expected that the leptonic Pontecorvo-Maki-Nakagawa-Sakata (PMNS) mixing matrix at the leading order is exactly compatible with a TBM
  \begin{eqnarray}
  \theta_{13}=0\,,\qquad \theta_{23}=\frac{\pi}{4}=45^\circ\,,\qquad\theta_{12}=\sin^{-1}\Big(\frac{1}{\sqrt{3}}\Big)\simeq35.3^\circ\,.
 \label{tbm}
 \end{eqnarray}
In order to explain the present terrestrial neutrino oscillation data, non-trivial next leading order corrections should be taken into account: for example, $(N^cN^c\Theta\Phi_{T})_{{\bf 1}}/\Lambda$, $(N^cN^c\Phi_{S}\Phi_{T})_{{\bf 1}}/\Lambda$, and $(LN^c\Phi_{T})_{{\bf 1}}H_u/\Lambda$.  
(For neutrino phenomenology we will consider in detail in the next project. See also an interesting paper\,\cite{Gehrlein:2017ryu}.).

Here the $U(1)_X$ quantum numbers associated to the charged-leptons are assigned in a way that (i) the charged lepton mass spectra are described and (ii) the ratio of electromagnetic $U(1)_X$-$[U(1)_{\rm EM}]^2$ and color anomaly $U(1)_X$-$[SU(3)_{C}]^2$ coefficients lies in the range\,\footnote{This range is derived from the bound ADMX experiment\,\cite{Asztalos:2003px} $(g_{a\gamma\gamma}/m_{a})^2\leq1.44\times10^{-19}\,{\rm GeV}^{-2}\,{\rm eV}^{-2}$.} $0<E/N<4$, where $E=\sum_f(\delta^{\rm G}_2X_{1f}+\delta^{\rm G}_1X_{2f})(Q^{\rm em}_{f})^2$ and $N=2\delta^{\rm G}_1\delta^{\rm G}_2$:
 \begin{eqnarray}
  \frac{E}{N}&=&\frac{23}{6}\,,\qquad\text{for}~{\cal Q}_{y_\tau}=-3q, ~{\cal Q}_{y_\mu}=-6q, ~{\cal Q}_{y_e}=11q\,;~\text{case-I} \label{cas1}\\
  \frac{E}{N}&=&\frac{1}{2}\,,~\,\qquad\text{for}~{\cal Q}_{y_\tau}=3q, \quad{\cal Q}_{y_\mu}=6q, ~{\cal Q}_{y_e}=-11q\,;~\text{case-II} \label{cas2}\\
 \frac{E}{N}&=&\frac{5}{2}\,,~\,\qquad\text{for}~{\cal Q}_{y_\tau}=3q, \quad{\cal Q}_{y_\mu}=6q, ~{\cal Q}_{y_e}=-11q\,;~\text{case-III}\,.
 \label{cas3}
 \end{eqnarray}
Similarly, the $U(1)_X$ quantum numbers associated to the neutrinos can be assigned by the anomaly-free condition of $U(1)_X$-$[gravity]^2$ together with the measured active neutrino observables:
 \begin{eqnarray}
 U(1)_X\times[gravity]^2&\propto&3\left\{4p-3q\right\}_{\rm quark}\nonumber\\
 &+&\left\{3p-{\cal Q}_{y^s_1}-{\cal Q}_{y^s_2}-{\cal Q}_{y^s_3}-{\cal Q}_{y_e}-{\cal Q}_{y_\mu}-{\cal Q}_{y_\tau}\right\}_{\rm lepton}=0\,.
 \label{ux_gr}
 \end{eqnarray}
 This vanishing anomaly, however, does not restrict ${\cal Q}_{y_{\nu}}$ (or equivalently ${\cal Q}_{y^{ss}_i}$), whose quantum  numbers can be constrained by the  new neutrino oscillations of astronomical-scale baseline,  as shown in Refs.\,\cite{Ahn:2018nfb, Ahn:2016hbn, Ahn:2016hhq}.
With the given above $U(1)_X$ quantum numbers, such $U(1)_X\times[gravity]^2$ anomaly is free for 
 \begin{eqnarray}
 21\,\frac{X_1}{2}=k_2\,X_2\qquad\text{with}~ k_2=\left\{\begin{array}{ll}
              11-\tilde{{\cal Q}}_{y^s_1}-\tilde{{\cal Q}}_{y^s_2}-\tilde{{\cal Q}}_{y^s_3};& \text{case-I}  \\
              1-\tilde{{\cal Q}}_{y^s_1}-\tilde{{\cal Q}}_{y^s_2}-\tilde{{\cal Q}}_{y^s_3};& \text{case-II}\\
              7-\tilde{{\cal Q}}_{y^s_1}-\tilde{{\cal Q}}_{y^s_2}-\tilde{{\cal Q}}_{y^s_3};& \text{case-III}
             \end{array}\right\}\,.
\label{cond1}
 \end{eqnarray}
 where $\tilde{{\cal Q}}_{y^s_i}={\cal Q}_{y^s_1}/X_2$. We choose $k_2=\pm21$ for the $U(1)_{X_i}$ charges to be smallest making no axionic domain-wall problem, as in Ref.\,\cite{Ahn:2018nfb, Ahn:2016hbn}. Hence, for the case-I $\tilde{{\cal Q}}_{y^s_1}+\tilde{{\cal Q}}_{y^s_2}+\tilde{{\cal Q}}_{y^s_3}=-10$ $(32)$; for the case-II $-20$ ($22$); for the case-III $-14$ ($28$), respectively, for $k_2=21(-21)$.
Then, the color anomaly coefficients are given by $\delta^{\rm G}_1=2X_1$ and $\delta^{\rm G}_2=-3X_2$, and subsequently the axionic domain-wall condition as in Ref.\,\cite{Ahn:2018nfb} is expressed with the reduced $k_1=\pm k_2=1$ as
 $N_1=4$ and $N_2=3$.
Clearly, in the QCD instanton backgrounds since the $N_1$ and $N_2$ are relative prime there is no $Z_{N_{\rm DW}}$
 discrete symmetry, and therefore no axionic domain-wall problem occurs.

Once the scalar fields $\Phi_{S}, \Theta, \tilde{\Theta},\Psi$ and $\tilde{\Psi}$ get VEVs, the flavor symmetry $U(1)_{X}\times SL_2(F_{3})$ is spontaneously broken\,\footnote{If the symmetry $U(1)_{X}$ is broken spontaneously, the massless modes $A_1$ of the scalar $\Phi_{S}$ (or $\Theta$) and $A_{2}$ of the scalar $\Psi(\tilde{\Psi})$ appear as phases.}. 
And at energies below the electroweak scale, all quarks and leptons obtain masses.
The relevant Yukawa interaction terms with chiral fermions $\psi$ charged under the flavored $ U(1)_X$ symmetry is given by 
 \begin{eqnarray}
  -{\cal L}_{YW} &=&
  \overline{q^{u}_{R}}\,\mathcal{M}_{u}\,q^{u}_{L}+\overline{q^{d}_{R}}\,\mathcal{M}_{d}\,q^{d}_{L}
   + \overline{\ell_{R}}\,{\cal M}_{\ell}\,\ell_{L}+\frac{g}{\sqrt{2}}W^+_\mu\overline{q^u_L}\gamma^\mu\,q^d_L+\frac{g}{\sqrt{2}}W^-_\mu\overline{\ell_L}\gamma^\mu\,\nu_L\nonumber\\
 &+& \frac{1}{2} \begin{pmatrix} \overline{\nu^c_L} & \overline{S_R} & \overline{N_R} \end{pmatrix} \begin{pmatrix} 0 & m^T_{DS} & m^T_D  \\ m_{DS} & e^{i\frac{A_{1}}{v_{\cal F}}}\,M_{S} & 0  \\ m_D & 0 & e^{i\frac{A_{1}}{v_{\cal F}}}\,M_R \end{pmatrix} \begin{pmatrix} \nu_L \\ S^c_R \\ N^c_R \end{pmatrix} +\text{h.c.}\,,
  \label{AxionLag2}
 \end{eqnarray}
where $g$ is the $SU(2)$ coupling constant, $q^{u}=(u,c,t)$, $q^{d}=(d,s,b)$, $\ell=(e, \mu, \tau)$, and $\nu_L=(\nu_e, \nu_\mu, \nu_\tau)$.
 
%%%%%%%%%%%%%%%%%%%%%%%%%%%%%%%%%%%%%%%%%%%%%%%%%%%%%%%%%%%%%%%%%%%%%%%%%%%%%%%
\subsubsection{Quarks and CKM mixings, and flavored-Axions}
\noindent
Now, let us move to discussion on the realization of quark masses and mixings, in which the physical mass hierarchies are directly responsible for the assignment of $U(1)_X$ quantum numbers.
The axion coupling matrices to the up- and down-type quarks, respectively, are diagonalized through bi-unitary transformations: $V^{\psi}_R{\cal M}_{\psi}V^{\psi\dag}_L=\hat{{\cal M}}_\psi$ (diagonal form), and the mass eigenstates $\psi'_R=V^\psi_R\,\psi_R$ and $\psi'_L=V^\psi_L\,\psi_L$. %These transformation include, in particular, the chiral transformation necessary to make ${\cal M}_u$ and ${\cal M}_d$ real and positive. This induces a contribution to the QCD vacuum angle. Note here that under the chiral rotation of the quark fields the effective QCD vacuum angle is invariant, see Refs.\,\cite{Ahn:2014gva, Ahn:2016hbn}.
With the desired VEV directions in Eq.\,(\ref{vev}), in the above Lagrangian\,(\ref{AxionLag2}) the mass matrices ${\cal M}_{u}$ and ${\cal M}_{d}$ for up- and down-type quarks, respectively, are expressed as
 \begin{eqnarray}
 &{\cal M}_{u}= {\left(\begin{array}{ccc}
 (iy_{u}\nabla_T-\tilde{y}_u\nabla^2_\eta)\nabla_\eta\,e^{i(\frac{A_1}{v_{\cal F}}-4\frac{A_{2}}{v_{g}})} &  0 &  0 \\
 0 &  (y_{c}+\frac{1-i}{2}\,\tilde{y}_c\nabla_T)\nabla_\eta\,e^{-2i\frac{A_{2}}{v_{g}}} &  0   \\
 0 &  0  &  \hat{y}_{t}
 \end{array}\right)}v_{u}\,,  \label{Ch1}\\
 &{\cal M}_{d}={\left(\begin{array}{ccc}
 (iy_d\nabla_S+\tilde{y}_{d}\nabla_T)\nabla_\eta\,e^{i(\frac{A_1}{v_{\cal F}}-\frac{A_{2}}{v_{g}})} &  0 &  0 \\
 \frac{1-i}{2}y_{d}\nabla_\eta\nabla_S\,e^{i(\frac{A_1}{v_{\cal F}}-\frac{A_{2}}{v_{g}})}  &  y_{s}\nabla_\eta\,e^{i\frac{A_{2}}{v_{g}}} &  0   \\
 3Y_{d}\nabla^2_S\,e^{i(5\frac{A_1}{v_{\cal F}}-\frac{A_{2}}{v_{g}})}  &  3Y_{s}\nabla^2_S\,e^{i(4\frac{A_1}{v_{\cal F}}+\frac{A_{2}}{v_{g}})}  & y_{b}\,e^{3i\frac{A_{2}}{v_{g}}}
 \end{array}\right)}v_{d}\,,
 \label{Ch2}
 \end{eqnarray}
where $v_{d}\equiv\langle H_{d}\rangle=v\cos\beta/\sqrt{2}$ and $v_{u}\equiv\langle H_{u}\rangle =v\sin\beta/\sqrt{2}$ with $v\simeq246$ GeV, and 
 \begin{eqnarray}
  \nabla_Q\equiv\frac{v_Q}{\sqrt{2}\,\Lambda} \quad\text{with}~Q=\eta, S, T, \Theta, \Psi, \tilde{\Psi}\,.
 \label{nabla}
 \end{eqnarray}
In the above mass matrices the corresponding Yukawa terms for up- and down-type quarks are given by
 \begin{eqnarray}
 y_{u}&=&\hat{y}_{u}\,\nabla_\Theta\nabla^4_{\tilde{\Psi}}\,,\qquad~ \tilde{y}_{u}=\hat{\tilde{y}}_{u}\,\nabla_\Theta\nabla^4_{\tilde{\Psi}}\,,\qquad~
 y_{c}=\hat{y}_{c}\,\nabla^2_{\tilde{\Psi}}\,,\qquad~
 \tilde{y}_{c}=\hat{\tilde{y}}_{c}\,\nabla^2_{\tilde{\Psi}}\,,\nonumber\\
 y_{d}&=& \hat{y}_{d}\,\nabla_{\tilde{\Psi}}\,,~~\quad\qquad~
 \tilde{y}_{d}=\hat{\tilde{y}}_{d}\,\nabla_{\tilde{\Psi}}\nabla_\Theta\,,\qquad~
 y_{s}=\hat{y}_{s}\,\nabla_\Psi\,,\qquad~ y_{b}=\hat{y}_{b}\,\nabla^3_\Psi\,.
 \label{Top1}
 \end{eqnarray}

%One of the most interesting features observed by experiments on the quarks is that the mass spectrum of the up-type quarks exhibits a much stronger hierarchical pattern to that of the down-type quarks, which may indicate that the CKM matrix\,\cite{PDG} is mainly generated by the mixing matrix of the down-type quark sector. Moreover, 
Due to the diagonal form of the up-type quark mass matrix in Eq.\,(\ref{ChL1}) the CKM mixing matrix $V_{\rm CKM}\equiv V^u_LV^{d\dag}_L$ coming from the charged quark-current term in Eq.\,(\ref{AxionLag2}) is generated from the down-type quark matrix in Eq.\,(\ref{Ch2}):
 \begin{eqnarray}
 V_{\rm CKM}=V^{d\dag}_L={\left(\begin{array}{ccc}
 1-\frac{1}{2}\lambda^2 & \lambda & A\lambda^3(\rho+i\eta)  \\
-\lambda & 1-\frac{1}{2}\lambda^2 & A\lambda^2   \\
 A\lambda^3(1-\rho+i\eta) & -A\lambda^2  & 1
 \end{array}\right)}+{\cal O}(\lambda^4)\,,
 \label{ckm0}
 \end{eqnarray}
in the Wolfenstein parametrization\,\cite{Wolfenstein:1983yz} and at higher precision\,\cite{Ahn:2011fg}, where $\lambda=0.22509^{+0.00091}_{-0.00071}$, $A=0.825^{+0.020}_{-0.037}$, $\bar{\rho}=\rho/(1-\lambda^2/2)=0.160^{+0.034}_{-0.021}$, and $\bar{\eta}=\eta/(1-\lambda^2/2)=0.350^{+0.024}_{-0.024}$ with $3\sigma$ errors\,\cite{ckm}.

The quark mass matrices ${\cal M}_{u}$ in Eq.\,(\ref{Ch1}) and ${\cal M}_{d}$ in Eq.\,(\ref{Ch2}) generate the up- and down-type quark masses:
 \begin{eqnarray}
 \widehat{\mathcal{M}}_{u}=P^{\ast}_{u}\,{\cal M}_{u}\,Q_{u}
 ={\rm diag}(m_{u},m_{c},m_{t})\,,\quad \widehat{\mathcal{M}}_{d}=V^{d\dag}_{R}\,{\cal M}_{d}\,V^{d}_{L}
 ={\rm diag}(m_{d},m_{s},m_{b})\,,
 \label{Quark21}
 \end{eqnarray}
where $P_u$ and $Q_u$ are diagonal phase matrices,  and $V^{d}_{L}$ and $V^{d}_{R}$ can be determined by diagonalizing the matrices for ${\cal M}^{\dag}_{d}{\cal M}_{d}$ and ${\cal M}_{d}{\cal M}^{\dag}_{d}$, respectively.
The physical structure of the up- and down-type quark Lagrangian should match up with the empirical up- and down-type quark masses and their ratios calculated from the measured PDG values\,\cite{PDG}:
 \begin{eqnarray}
 \frac{m_{d}}{m_{b}}&\doteqdot&1.12^{+0.13}_{-0.11}\times10^{-3}\,,\qquad \frac{m_{s}}{m_{b}}\doteqdot2.30^{+0.21}_{-0.12}\times10^{-2}\,,\qquad\frac{m_{u}}{m_{t}} \doteqdot2.41^{+0.03}_{-0.03}\times10^{-2}\,,
\nonumber\\
  \frac{m_{u}}{m_{d}} &\doteqdot&0.38-0.58\,,\,\qquad\qquad\frac{m_{c}}{m_{t}}\doteqdot7.39^{+0.20}_{-0.20}\times10^{-3}\,,\qquad
  \frac{m_{u}}{m_{c}}\doteqdot1.72^{+0.52}_{-0.34}\times10^{-3}\,, \label{mQRatio}\\
  m_b&=&4.18^{+0.04}_{-0.03}\,{\rm GeV}\,,\qquad\quad~ m_c=1.28\pm0.03\,{\rm GeV}\,,\qquad~ m_t=173.1\pm0.6\,{\rm GeV}\,,
\label{qumas}
 \end{eqnarray}
where $c$- and $b$-quark masses are the running masses in the $\overline{\rm MS}$ scheme, and the light $u$-, $d$-, $s$-quark masses are the current quark masses in the $\overline{\rm MS}$ scheme at the momentum scale $\mu\approx2$ GeV.
So, the following new expansion parameters are defined in a way that the diagonalizing matrix $V^{d}_L$ satisfies the CKM matrix as well as the empirical quark masses and their ratios in Eqs.\,(\ref{mQRatio}) and (\ref{qumas}):
\begin{eqnarray}
\nabla_T&=&\kappa\frac{|\hat{y}_d|}{|\hat{\tilde{y}}_d|}\qquad\text{with}~\phi_{\tilde{d}}=-\phi_d-\frac{\pi}{2}\,,\label{mdi}\\
 \nabla_\Theta&=&\frac{1}{\kappa}\nabla_S=\Big|\frac{X_2\delta^{\rm G}_1}{X_1\delta^{\rm G}_2}\Big|\sqrt{\frac{2}{1+\kappa^2}}\nabla_\Psi\,,
  \label{expan_1}\\
 \nabla_\eta&=&\Big(\frac{m_c}{m_t}\Big)_{\rm PDG}\Big|\frac{\hat{y}_t}{\hat{y}_c+\hat{\tilde{y}}_c\nabla_T}\Big|\frac{1}{\nabla^2_\Psi}\,,\label{expan_2}\\
 \nabla_\Psi&\simeq&\lambda\Big|\frac{X_1\delta^{\rm G}_2}{X_2\delta^{\rm G}_1}\Big|^{\frac{2}{3}}\Big(\frac{B\,(1+\kappa^2)}{6\kappa^2}\frac{|\hat{y}_b|}{|\hat{Y}_{d1}+3\kappa^2\hat{Y}_{d2}|}\Big)^{\frac{1}{3}}\,.\label{expan_0}
\end{eqnarray}
 Then, the mixing matrix $V^{d\dag}_{L}=V_{\rm CKM}$ is obtained by diagonalizing the Hermitian matrix ${\cal M}^\dag_{d}{\cal M}_{d}$:
 \begin{eqnarray}
 V^{d}_{L}{\cal M}^\dag_{d}{\cal M}_{d}V^{d\dag}_{L}={\rm diag}(|m_{d}|^{2}, |m_s|^{2}, |m_{b}|^{2})\,.
 \label{MDMD0}
 \end{eqnarray}
The CKM mixing angles in the standard parametrization\,\cite{Chau:1984fp} can be roughly described as
 \begin{eqnarray}
\theta^q_{12}\simeq \frac{1}{\sqrt{2}}\Big|\frac{\hat{y}_d}{\hat{y}_s}\Big|\,\nabla_S\,,\quad \theta^q_{23}\simeq 3\kappa^2\Big|\frac{\hat{Y}_{s1}+3\kappa^2\hat{Y}_{s2}}{\hat{y}_b}\Big|\frac{\nabla^4_\Theta}{\nabla_\Psi^2}\,,\quad\theta^q_{13}\simeq 3\Big|\frac{\hat{Y}_{d1}+3\kappa^2\hat{Y}_{d2}}{\hat{y}_b}\Big|\,\nabla_\Psi\nabla^2_S\,.
 \end{eqnarray}
And with the quark fields redefinition the CKM CP phase is given as
 \begin{eqnarray}
 \delta^{q}_{CP}\equiv\tan^{-1}\left(\eta/\rho\right)=\phi^{d}_{2}-2\phi^{d}_{3}\,,
 \end{eqnarray}
 where $\phi^{d}_{2}\simeq\arg\{(\hat{Y}^\ast_{d1}+3\kappa^2\hat{Y}^\ast_{d2})\hat{y}_b\}/2-\phi^{d}_{1}/2$ and $2\phi^{d}_{3}\simeq\arg(\hat{y}^\ast_s\hat{y}_b)+\phi^{d}_{1}-\phi^{d}_{2}+\pi/4$, and $\phi^{d}_{1}=\arg\{(\hat{Y}^\ast_{s1}+3\kappa^2\hat{Y}^\ast_{s2})\hat{y}_b\}/2$.  As designed, the CKM matrix is well described with $J^{\rm quark}_{CP}={\rm Im}[V_{us}V_{cb}V^{\ast}_{ub}V^{\ast}_{cs}]\simeq A^{2}\lambda^{6}\sqrt{\rho^{2}+\eta^{2}}\sin\delta^{q}_{CP}$.
Subsequently, the up- and down-type quark masses are obtained as
 \begin{eqnarray}
 m_t&\simeq&|\hat{y}_t|\,v_u\,, \qquad\qquad\qquad\qquad\qquad\qquad\qquad m_b\simeq|\hat{y}_b|\nabla^3_\Psi\,v_b\,,\nonumber\\
 m_c&\simeq&\Big|\hat{y}_c+\frac{1-i}{2}\hat{\tilde{y}}_c\nabla_T\Big|\nabla^2_\Psi\nabla_\eta\,v_u\,, \,\qquad\qquad\quad\, m_s\simeq|\hat{y}_s|\nabla_\Psi\nabla_\eta\,v_d\,,\nonumber\\
 m_u&\simeq&\nabla^4_\Psi\nabla_\eta\nabla_\Theta|i\hat{y}_u\nabla_T-\hat{\tilde{y}}_u\nabla^2_\eta|\,v_u\,,\qquad\qquad\,~ m_d\simeq2|\hat{y}_d\sin\phi_d|\nabla_\Psi\nabla_\eta\nabla_S\,v_d\,.
 \label{qm}
 \end{eqnarray}
And the parameter of $\tan\beta\equiv v_u/v_d$ is given in terms of the PDG value in Eq.\,(\ref{qumas}) by
 \begin{eqnarray}
 \tan\beta\simeq\Big(\frac{m_t}{m_b}\Big)_{\rm PDG}\Big|\frac{\hat{y}_b}{\hat{y}_t}\Big|\,\nabla^3_\Psi\,.
 \label{}
 \end{eqnarray}

Since all the parameters in the quark sector are correlated with one another, it is very crucial for obtaining the values of the new expansion parameters to reproduce the empirical results of the CKM mixing angles and quark masses. Moreover, since such parameters are also closely correlated with those in the lepton sector, finding the value of parameters is crucial to produce the empirical results of the charged leptons (see below Eq.\,(\ref{ChL1})) and the light active neutrino masses in our model.
%%%%%%%%%%%%%%%%%%%%%%%%%%%%%%%%%%%%%%%%%%%%%%%%%%%%%%%%%%%%%%%%%%%%%%
\subsubsection{Numerical analysis for Quark masses and CKM mixing angles}
\label{num_quark}
We perform a numerical simulation\,\footnote{Here, in numerical calculation, we have only considered the mass matrices in Eqs.\,(\ref{Ch1}) and (\ref{Ch2}) since it is expected that the corrections to the VEVs due to higher dimensional operators could be small enough below a few percent level.} using the linear algebra tools of Ref.\,\cite{Antusch:2005gp}. With the inputs 
\begin{eqnarray}
\tan\beta=4.7\,, \qquad\kappa=0.33\,,
  \label{tanpara}
\end{eqnarray}
and $|\hat{y}_d|=1.1$ ($\phi_d=3.070$ rad), $|\hat{\tilde{y}}_d|=1.194$, $|\hat{y}_s|=0.370$ ($\phi_s=4.920$ rad), $|\hat{y}_b|=2.280$ ($\phi_b=0$), $|\hat{y}_u|=0.400$ ($\phi_u=0$), $|\hat{\tilde{y}}_u|=1.0$ ($\phi_{\tilde{u}}=0$), $|\hat{y}_c|=2.800$ ($\phi_c=3.600$ rad), $|\hat{\tilde{y}}_c|=1.000$ ($\phi_{\tilde{c}}=0$), $|\hat{y}_t|=1.017$ ($\phi_t=0$), $|\hat{Y}_{d1}|=0.900$ ($\phi_{Y_{d1}}=4.800$ rad), $|\hat{Y}_{d2}|=0.800$ ($\phi_{Y_{d2}}=0$), $|\hat{Y}_{s1}|=2.600$ ($\phi_{Y_{s1}}=6.500$ rad), $|\hat{Y}_{s2}|=1.900$ ($\phi_{Y_{s2}}=0.117$ rad), leading to 
\begin{eqnarray}
\nabla_\Psi=0.370\,, \quad\nabla_S=0.109\,, \quad\nabla_T=0.304\,, \quad \nabla_\eta=0.020\,,
 \label{quarkvalue}
 \end{eqnarray} 
we obtain the mixing angles and Dirac CP phase $\theta^q_{12}=12.98^{\circ}$, $\theta^q_{23}=2.32^{\circ}$, $\theta^q_{13}=0.22^{\circ}$, $\delta^q_{CP}=65.18^{\circ}$ compatible with the $3\sigma$ Global fit of CKMfitter\,\cite{ckm}; the quark masses $m_d=4.49$ MeV, $m_s=101.62$ MeV, $m_b=4.18$ GeV, $m_u=2.57$ MeV, $m_c=1.28$ GeV, and $m_t=173.1$ GeV compatible with the values in PDG\,\cite{PDG}.
 
 %%%%%%%%%%%%%%%%%%%%%%%%%%%%%%%%%%%%%%%%%%%%%%%%%%%%%%%%%%%%%%%%%%%%%%%%%%%
\subsubsection{charged-Leptons and flavored-Axions}
\noindent
According to the $U(1)_X$ charge assignment of the charged-leptons in Eqs.\,(\ref{cas1}), (\ref{cas2}), and (\ref{cas3}), the charged-lepton mass matrix in the Lagrangian (\ref{AxionLag2}) is written as
 \begin{eqnarray}
 {\cal M}_{\ell}&=& {\left(\begin{array}{ccc}
 y_{e}\,e^{{\cal Q}_{e}i\frac{A_{2}}{v_{g}}} & 0 &  0 \\
 0 & y_{\mu}\,e^{{\cal Q}_{\mu}i\frac{A_{2}}{v_{g}}} & 0 \\
 0 & 0 & y_{\tau}\,e^{{\cal Q}_{\tau}i\frac{A_{2}}{v_{g}}}
 \end{array}\right)}v_{d}\,,
 \label{ChL1}
 \end{eqnarray}
where ${\cal Q}_{e}=-11$, ${\cal Q}_{\mu}=6$, ${\cal Q}_{\tau}=3$ for the case-I ($E/N=23/6$); ${\cal Q}_{e}=11$, ${\cal Q}_{\mu}=-6$, ${\cal Q}_{\tau}=-3$ for the case-II ($E/N=1/2$); ${\cal Q}_{e}=11$, ${\cal Q}_{\mu}=-6$, ${\cal Q}_{\tau}=-3$ for the case-III ($E/N=5/2$). 
The corresponding Yukawa terms are expressed in terms of Eqs.\,(\ref{nabla}) and (\ref{expan_0}) used in the quark sector as
 \begin{eqnarray}
 y_{e}&=&\hat{y}_{e}\,\nabla^{11}_{\Psi}\,,\qquad\qquad y_{\mu}=\hat{y}_{\mu}\,\nabla^{6}_{\Psi}\,,\qquad\qquad
 y_{\tau}=\hat{y}_{\tau}\,\nabla^3_{\Psi}\,,
 \label{cLep1}
 \end{eqnarray}
where $\nabla_\Psi=\nabla_{\tilde{\Psi}}$ in SUSY limit  is used. And the hat Yukawa couplings $\hat{y}_{e,\mu,\tau}$ are fixed\,\footnote{The charged lepton sector, in common with the quark sector, has VEV corrections and the hat Yukawa couplings are corrected.} as $\hat{y}_{e}=0.793152$, $\hat{y}_{\mu}=1.137250$, $\hat{y}_{\tau}=0.968747$ by using the numerical values of Eq.\,(\ref{quarkvalue}) in quark sector via the empirical results $m_e=0.511$ MeV, $m_\mu=105.683$ MeV, and $m_{\tau}=1776.86$ MeV\,\cite{PDG}.

%%%%%%%%%%%%%%%%%%%%%%%%%%%%%%%%%%%%%%%%%%%%%%%%%%%%%%%%%%%%%%%%%%%%%%%%%%%%%%%
\section{Scale of PQ phase transition and QCD axion properties}
\noindent The couplings of the flavored-axions and the mass of the QCD axion are inversely proportional to the PQ symmetry breaking scale. In a theoretical view of Refs.\cite{Ahn:2014gva, Ahn:2016hbn, Ahn:2018nfb}, the scale of PQ symmetry breakdown congruent to that of the seesaw mechanism can push the scale much beyond the electroweak scale, rendering the flavored-axions very weakly interacting particles. 
Since the weakly coupled flavored-axions (one linear combination QCD axion and its orthogonal ALP) could carry away a large amount of energy from the interior of stars, according to the standard stellar evolution scenario their couplings should be bounded with electrons\,\footnote{The second ($\mu$) and third ($\tau$) generation particles are absent in almost all astrophysical objects.}, photons, and nucleons. Hence, such weakly coupled flavored-axions have a wealth of interesting phenomenological implications in the context of astro-particle physics\,\cite{Ahn:2016hbn, Ahn:2018nfb}, like the formation of a cosmic diffuse background of axions from the Sun\,\cite{Aprile:2014eoa, Fu:2017lfc}; from evolved low-mass stars, such as red-giants and horizontal-branch stars in globular clusters\,\cite{Redondo:2013wwa, Viaux:2013lha}, or white dwarfs\,\cite{Raffelt:1985nj, Bertolami:2014wua}; from neutron stars\,\cite{Keller:2012yr}; and from the duration of the neutrino burst of the core-collapse supernova SN1987A\,\cite{Raffelt:2006cw} as well as the rare flavor changing decay processes induced by the flavored-axions $K^+\rightarrow\pi^++A_i$\,\cite{Adler:2008zza, Wilczek:1982rv} and $\mu\rightarrow e+\gamma+A_i$\,\cite{Bolton:1988af, Wilczek:1982rv}  etc..

Such flavored-axions could be produced in hot astrophysical plasmas, thus transporting energy out of stars and other astrophysical objects, and they could also be produced by the rare flavor changing decay processes.
Actually, the coupling strength of these particles with normal matter and radiation is bounded by the constraint that stellar lifetimes and energy-loss rates\,\cite{raffelt} as well as the branching ratios for the $\mu$ and $K$ flavor changing decays\,\cite{Adler:2008zza, Bolton:1988af} should not be counter to observations.
Interestingly enough, the recent observations also show a preference for extra energy losses in stars at different evolutionary stages - red giants, supergiants, helium core burning stars, white dwarfs, and neutron stars (see Ref.\,\cite{Giannotti:2017hny} for the summary of extra cooling observations and Ref.\,\cite{Ahn:2016hbn} on the interpretation to a bound of the QCD axion decay constant); the present experimental limit, ${\rm Br}(K^+\rightarrow\pi^+A_i)<7.3\times10^{-11}$\,\cite{Adler:2008zza}, puts a lower bound on the axion decay constant, and in the near future the NA62 experiment expected to reach the sensitivity ${\rm Br}(K^+\rightarrow\pi^+A_i)<1.0\times10^{-12}$\,\cite{Fantechi:2014hqa} will probe the flavored-axions or put a severe bound on the QCD axion decay constant $F_A$ (or flavored-axion decay constants $F_{a_i}=f_{a_i}/\delta^{\rm G}_i$).
According to the recent investigation in Ref.\,\cite{Ahn:2016hbn, Ahn:2018nfb}, the flavored-axions (QCD axion and its orthogonal ALP) would provide very good hints for a new physics model for quarks and leptons. Fortunately, in a framework of the flavored-PQ symmetry the cooling anomalies hint at an axion coupling to electrons, photons, and neutrons, which should not conflict with the current upper bound on the rare $K^+\rightarrow \pi^+ A_i$ decay. Remark that once a scale of PQ symmetry breakdown is fixed the other is automatic including the QCD axion decay constant and the mass scale of heavy neutrino associated to the seesaw mechanism.

In order to fix the QCD axion decay constant $F_A$ (or flavored-axion decay constants $F_{a_i}=f_{a_i}/\delta^{\rm G}_i$), we will consider two tight constraints coming from astro-particle physics: axion cooling of stars via bremsstrahlung off electrons and flavor-violating processes induced by the flavored-axions. 
 %%%%%%%%%%%%%%%%%%%%%%%%%%%%%%%%%
\subsection{Flavored-Axion cooling of stars via bremsstrahlung off electrons}
 \noindent %In the so-called flavored-axion framework, generically, the SM charged lepton fields are nontrivially $U(1)_X$-charged Dirac fermions, and thereby the flavored-axion coupling to electrons are added to the Lagrangian through a chiral rotation.
 
In the present model since the flavored-axion $A_2$ couples directly to electrons, the axion can be emitted by Compton scattering, atomic axio-recombination and axio-deexcitation, and axio-bremsstrahlung in electron-ion or electron-electron collision\,\cite{Redondo:2013wwa}.
The flavored-axion $A_2$ coupling to electrons in the model reads 
\begin{eqnarray}
 g_{Aee}=\frac{X_e\,m_e}{\sqrt{2}\,\delta^{\rm G}_2\,F_A}
\end{eqnarray}
where $m_e=0.511$ MeV, $F_A=f_{a_i}/\sqrt{2}\delta^{\rm G}_i$, $\delta^{\rm G}_2=-3X_2$ and $X_e=-11X_2$.
 Indeed, the longstanding anomaly in the cooling of WDs (white dwarfs) and RGB (red giants branch) stars in globular clusters where bremsstrahlung off electrons is mainly efficient\,\cite{Raffelt:1985nj} could be explained by axions with the fine-structure constant of axion to electrons $\alpha_{Aee}=(0.29-2.30)\times10^{-27}$\,\cite{WD01} and  $\alpha_{Aee}=(0.41-3.70)\times10^{-27}$\,\cite{Bertolami:2014wua, wd_recent2}, indicating the clear systematic tendency of stars to cool faster than  predicted. It is recently reexamined in Ref.\,\cite{Giannotti:2017hny} as Eq.\,(\ref{AP_cons})
%\begin{eqnarray}
% 6.7\times10^{-29}\lesssim\alpha_{Aee}\lesssim5.6\times10^{-27}\quad\text{at}~3\sigma\,,
%\end{eqnarray}
where $\alpha_{Aee}=g^2_{Aee}/4\pi$, which is interpreted in terms of the QCD axion decay constant in the present model as 
\begin{eqnarray}
 0.5\times10^{10}\lesssim F_{A}[{\rm GeV}]\lesssim4.4\times10^{10}\,.
 \label{cons_1}
\end{eqnarray}
This bound comes from the $U(1)_X$ quantum number of electron, $X_e=-11X_2$, as shown in Eq.\,(\ref{ChL1}). Note that the $U(1)_X$ quantum number of charged leptons in Eqs.\,(\ref{cas1}, \ref{cas2}, \ref{cas3}) is different from the one in Ref.\,\cite{Ahn:2018nfb} because of the different flavor structures of Yukawa interactions, see Eqs.\,(\ref{Ch1}), (\ref{Ch2}), and (\ref{ChL1}), leading to the different values of expansion parameters Eqs.\,(\ref{tanpara}) and (\ref{quarkvalue}) satisfying the empirical quark and lepton masses and mixings.

 %%%%%%%%%%%%%%%%%%%%%%%%%%%%
\subsection{Flavor-Changing process $K^+\rightarrow\pi^++A_i$ induced by the flavored-axions}
\noindent Below the QCD scale (1 GeV$\approx4\pi f_\pi$), the chiral symmetry is broken and $\pi$ and $K$, and $\eta$ are produced as pseudo-Goldstone bosons.
Since a direct interaction of the SM gauge singlet flavon fields charged under $U(1)_X$ %responsible for spontaneous symmetry breaking 
with the SM quarks charged under $U(1)_X$ can arise through Yukawa interaction, the flavor-changing process $K^+\rightarrow\pi^++A_i$ is induced by the flavored-axions $A_i$.
Then, the flavored-axion interactions with the flavor violating coupling to the $s$- and $d$-quark is given by
\begin{eqnarray}
 {\cal L}^{A_isd}_Y\simeq i\Big(\frac{|X_1|\,A_1}{2f_{a_1}}-\frac{|X_2|\,A_2}{f_{a_2}}\Big)\bar{s}d\,(m_s-m_d)\lambda\Big(1-\frac{\lambda^2}{2}\Big)\,,
  \label{}
\end{eqnarray}
where\,\footnote{In the standard parametrization the mixing elements of $V^d_R$ are given by $\theta^R_{23}\simeq A\lambda^2(\nabla_\eta/\kappa^2\nabla^2_\Psi)\,|\hat{y}_s/\hat{y}_b|$, $\theta^R_{13}\simeq AB\lambda^5|\sin\phi_d|\,|\hat{y}_d/(\hat{Y}_{s1}+3\kappa^2\hat{Y}_{s2})|\,(2\nabla_\eta/3\kappa\nabla^3_\Psi)$, and $\theta^R_{12}\simeq2\sqrt{2}|\sin\phi_d|\lambda^2$. Its effect to the flavor violating coupling to the $s$- and $d$-quark is negligible: $(V^d_R\,{\rm Diag.}(-4\frac{A_1}{v_{\cal F}}, -4\frac{A_1}{v_{\cal F}}, 0)\,V^{d\dag}_R)_{12}=0$ at leading order.} $V^{d\dag}_L= V_{\rm CKM}$, $f_{a_1}=|X_1|v_{\Theta}(1+\kappa^2)^{1/2}$, and $f_{a_2}=|X_2|v_g$ are used. Then the decay width of $K^+\rightarrow\pi^++A_i$ is given by\,\cite{Wilczek:1982rv, raredecay}
 \begin{eqnarray}
   \Gamma(K^+\rightarrow\pi^++A_i)=\frac{m^3_K}{16\pi}\Big(1-\frac{m^2_{\pi}}{m^2_{K}}\Big)^3\big|{\cal M}_{dsi}\big|^2\,,
  \label{}
 \end{eqnarray}
 where $m_{K^{\pm}}=493.677\pm0.013$ MeV, $m_{\pi^{\pm}}=139.57018(35)$ MeV\,\cite{pdg17}, and
 \begin{eqnarray}
   \big|{\cal M}_{ds1}\big|^2=\Big|\frac{X_1}{2\sqrt{2}\delta^{\rm G}_1\,F_{A}}\lambda\Big(1-\frac{\lambda^2}{2}\Big)\Big|^2\,,\qquad \big|{\cal M}_{ds2}\big|^2=\Big|\frac{X_2}{\sqrt{2}\delta^{\rm G}_2\,F_{A}}\lambda\Big(1-\frac{\lambda^2}{2}\Big)\Big|^2\,,
  \label{}
 \end{eqnarray}
where $F_A=f_{a_i}/(\delta^{\rm G}_i\sqrt{2})$ is used.
From the present experimental upper bound in Eq.\,(\ref{AP_cons}), ${\rm Br}(K^+\rightarrow\pi^+A_i)<7.3\times10^{-11}$, with ${\rm Br}(K^+\rightarrow\pi^+\nu\bar{\nu})=1.73^{+1.15}_{-1.05}\times10^{-10}$\,\cite{Artamonov:2008qb},
we obtain the lower limit on the QCD axion decay constant
 \begin{eqnarray}
              F_{A}\gtrsim2.72\times10^{10}\,{\rm GeV}\,.
 \label{cons_2}
 \end{eqnarray}
Hence, from Eqs.\,(\ref{cons_1}) and (\ref{cons_2}) we can obtain a strongest bound on the QCD axion decay constant 
 \begin{eqnarray}
 F_{A}=3.56^{\,+0.84}_{\,-0.84}\times10^{10}\,{\rm GeV}\,.
  \label{k_bound}
 \end{eqnarray}
Interestingly enough, from Eqs.\,(\ref{quarkvalue}) and (\ref{k_bound}) the scale $\Lambda=3F_A/(\sqrt{2}\,\nabla_\Psi)$ responsible for the FN mechanism can be determined   
 \begin{eqnarray}
 \Lambda=2.04^{\,+0.48}_{\,-0.48}\times10^{11}\,{\rm GeV}\,.
  \label{lamScale}
 \end{eqnarray}
 
In the near future the NA62 experiment will be expected to reach the sensitivity of ${\rm Br}(K^+\rightarrow\pi^++A_i)<1.0\times10^{-12}$\,\cite{Fantechi:2014hqa}, which is interpreted as the flavored-axion decay constant and its corresponding QCD axion decay constant 
 \begin{eqnarray}
 f_{a_i}>9.86\times10^{11}\,{\rm GeV}\,\Leftrightarrow\, F_{A}>2.32\times10^{11}\,{\rm GeV}\,.
  \label{knew_bound}
 \end{eqnarray}
Clearly, the NA62 experiment will probe the flavored-axions or exclude the present model.

%%%%%%%%%%%%%%%%%%%%%%%%%
\subsection{QCD axion interactions with nucleons}
\noindent Below the chiral symmetry breaking scale, the axion-hadron interactions are meaningful (rather than the axion-quark interactions) for the axion production rate in the core of a star where the temperature is not as high as $1$ GeV, which is given by\,\cite{Ahn:2014gva}
 \begin{eqnarray}
  -{\cal L}^{a-\psi_N} &=& \frac{\partial_{\mu}a}{2F_{A}}\,X_{\psi_N}\overline{\psi}_N\,\gamma_\mu\gamma^5\,\psi_N
  \label{a_nucleon}
 \end{eqnarray}
where $a$ is the QCD axion, its decay constant is given by $F_A=f_A/N$ with $f_A=\sqrt{2}\,\delta^{\rm G}_2f_{a_1}=\sqrt{2}\,\delta^{\rm G}_1f_{a_2}$, and $\psi_N$ is the nucleon doublet $(p,n)^T$ (here $p$ and $n$ correspond to the proton field and neutron field, respectively). Recently, the couplings of the axion to the nucleon are very precisely extracted as\,\cite{diCortona:2015ldu}
 \begin{eqnarray}
  X_p&=&-0.47(3)+0.88(3)\frac{\tilde{X}_u}{N}-0.39(2)\frac{\tilde{X}_d}{N}-0.038(5)\frac{\tilde{X}_s}{N}\nonumber\\
  &&-0.012(5)\frac{\tilde{X}_c}{N}-0.009(2)\frac{\tilde{X}_b}{N}-0.0035(4)\frac{\tilde{X}_t}{N}\,,\\
  X_n&=&-0.02(3)+0.88(3)\frac{\tilde{X}_d}{N}-0.39(2)\frac{\tilde{X}_u}{N}-0.038(5)\frac{\tilde{X}_s}{N}\nonumber\\
  &&-0.012(5)\frac{\tilde{X}_c}{N}-0.009(2)\frac{\tilde{X}_b}{N}-0.0035(4)\frac{\tilde{X}_t}{N}\,,
 \label{Nucleon_Lagran01}
 \end{eqnarray}
 where $N=2\delta^{\rm G}_{1}\delta^{\rm G}_{2}$ with $\delta^{\rm G}_{1}=2X_1$ and $\delta^{\rm G}_{2}=-3X_2$, and $\tilde{X}_q=\delta^{\rm G}_{2}X_{1q}+\delta^{\rm G}_{1}X_{2q}$ with $q=u,d,s$ and $X_{1u}=X_1$, $X_{1d}=X_1$, $X_{1s}=0$, $X_{1c}=0$, $X_{1b}=0$, $X_{1t}=0$, $X_{2u}=-4X_2$, $X_{2d}=-X_2$, $X_{2s}=X_2$, $X_{2c}=-2X_2$, $X_{2b}=3X_2$, $X_{2t}=0$. 
And the QCD axion coupling to the neutron is written as
 \begin{eqnarray}
  g_{Ann}=\frac{|X_n|\,m_n}{F_A}\,,
  \label{coupling_n01}
 \end{eqnarray}
where the neutron mass $m_n=939.6$ MeV. The state-of-the-art upper limit on this coupling, $g_{Ann}<8\times10^{-10}$\,\cite{Sedrakian:2015krq}, from the neutron star cooling is interpreted as the lower bound of the QCD axion decay constant
 \begin{eqnarray}
   F_A>9.53\times10^{7}\,{\rm GeV}\,.
  \label{a_nucleon01}
 \end{eqnarray}
Clearly, the strongest bound on the QCD axion decay constant comes from the flavored-axion cooling of stars via bremsstrahlung off electrons in Eq.\,(\ref{cons_1}) as well as the flavor-changing process $K^+\rightarrow\pi^++A_i$ induced by the flavored-axions in Eq.\,(\ref{cons_2}). 

Using the state-of-the-art calculation in Eq.\,(\ref{Nucleon_Lagran01}) and the QCD axion decay constant in Eq.\,(\ref{k_bound}), we can obtain 
 \begin{eqnarray}
  g_{Ann}=2.14^{+0.66}_{-0.41}\times10^{-12}\,,
  \label{}
 \end{eqnarray}
which is {\it incompatible} with the hint for extra cooling from the neutron star in the supernova remnant ``Cassiopeia A" by axion neutron bremsstrahlung, $g_{Ann}=3.74^{+0.62}_{-0.74}\times10^{-10}$\,\cite{Leinson:2014ioa}. This huge discrepancy may be explained by considering other means in the cooling of the superfluid core in the neutron star, for example, by neutrino emission in pair formation in a multicomponent superfluid state $^3{\rm P}_2\,(m_j=0,\pm1, \pm2)$\,\cite{Leinson:2014cja}.

%%%%%%%%%%%%%%%%%%%%%%%%%%%%%%%%
\subsection{QCD axion mass and its interactions with photons}
\noindent With the well constrained QCD axion decay constant in Eq.\,(\ref{k_bound}) congruent to the seesaw scale we can predict the QCD axion mass and its corresponding axion-photon coupling.

As in Refs.\,\cite{Ahn:2014gva, Ahn:2016hbn}, the axion mass in terms of the pion mass and pion decay constant is obtained as
 \begin{eqnarray}
 m^{2}_{a}F^{2}_{A}=m^{2}_{\pi^0}f^{2}_{\pi}F(z,w)\,,
\label{axiMass2}
 \end{eqnarray}
where\,\footnote{Here $F(z,\omega)$ can be replaced in high accuracy as in Ref.\,\cite{diCortona:2015ldu} by $F(z)=\frac{z}{(1+z)^2}\big\{1+2\frac{m^2_{\pi^0}}{f^2_\pi}\big(h_r+\frac{z^2-6z+1}{(1+z)^2}\,l_r\big)\big\}$, where $h_r=(4.8\pm1.4)\times10^{-3}$ and $l_r=7(4)\times10^{-3}$.} $f_\pi=92.21(14)$ MeV\,\cite{PDG} and
 \begin{eqnarray}
 F(z,w)=\frac{z}{(1+z)(1+z+w)}\quad\text{with}~z\equiv\frac{m^{\overline{\rm MS}}_u(2{\rm GeV})}{m^{\overline{\rm MS}}_d(2{\rm GeV})}=0.48(3)~\text{and}~\omega=0.315\,z\,.
 \label{axipra}
 \end{eqnarray}
Note that the Weinberg value lies in $0.38<z<0.58$\,\cite{PDG, Leutwyler:1996qg}.
After integrating out the heavy $\pi^{0}$ and $\eta$ at low energies, there is an effective low energy Lagrangian with an axion-photon coupling $g_{a\gamma\gamma}$:
 \begin{eqnarray}
{\cal L}_{a\gamma\gamma}= \frac{1}{4}g_{a\gamma\gamma}\,a_{\rm phys}\,F^{\mu\nu}\tilde{F}_{\mu\nu}=-g_{a\gamma\gamma}\,a_{\rm phys}\,\vec{E}\cdot\vec{B}\,,
 \end{eqnarray}
where $\vec{E}$ and $\vec{B}$ are the electromagnetic field components.
And the axion-photon coupling can be expressed in terms of the QCD axion mass, pion mass, pion decay constant, $z$ and $w$:
 \begin{eqnarray}
 g_{a\gamma\gamma}=\frac{\alpha_{\rm em}}{2\pi}\frac{m_a}{f_{\pi}m_{\pi^0}}\frac{1}{\sqrt{F(z,w)}}\left(\frac{E}{N}-\frac{2}{3}\,\frac{4+z+w}{1+z+w}\right)\,.
 \label{gagg}
 \end{eqnarray}
The upper bound on the axion-photon coupling is derived from the recent analysis of the horizontal branch (HB) stars in galactic globular clusters (GCs)\,\cite{Ayala:2014pea}, which translates into the lower bound of decay constant through Eq.\,(\ref{axiMass2}), as
  \begin{eqnarray}
 |g_{a\gamma\gamma}|<6.6\times10^{-11}\,{\rm GeV}^{-1}\,(95\%\,{\rm CL})\,\Leftrightarrow F_A\gtrsim\left\{
       \begin{array}{lll}
         3.23\times10^{7}\,{\rm GeV}\, & \hbox{case-I} \\
         2.64\times10^{7}\,{\rm GeV}, & \hbox{case-II}\\
         8.84\times10^{6}\,{\rm GeV}, & \hbox{case-III}
       \end{array}
     \right.
   \label{gagg_const}
 \end{eqnarray}
where in the right side $E/N=23/6$ (case-I), $1/2$ (case-II), $5/2$ (case-III) for $z=0.48$ are used. Subsequently, the bound in Eq.\,(\ref{gagg_const}) translates into the upper bound of axion mass through Eq.\,(\ref{axiMass2}) as $m_a\lesssim0.17$ eV, $\lesssim0.21$ eV, and $\lesssim0.62$ eV for the case-I, -II, and -III, respectively. It is well know that magnetic fields in or behind galaxy clusters convert photons into axions and alter the spectrum of the X-ray photons arriving at the earth\,\cite{Sikivie:1983ip, Raffelt:1987im}. The non-observation of the X-ray spectral modulations induced by axion to photon conversion with data drawn from the Chandra archive has placed a bound on the axion-photon coupling\,\cite{Conlon:2017qcw}
  \begin{eqnarray}
 |g_{a\gamma\gamma}|\lesssim1.5\times10^{-12}\,{\rm GeV}^{-1}\,(95\%\,{\rm CL})\,\Leftrightarrow F_A\gtrsim\left\{
       \begin{array}{lll}
         1.42\times10^{9}\,{\rm GeV}\, & \hbox{case-I} \\
         1.16\times10^{9}\,{\rm GeV}, & \hbox{case-II}\\
         3.89\times10^{8}\,{\rm GeV}, & \hbox{case-III}
       \end{array}
     \right.\,.
   \label{gagg_constN}
 \end{eqnarray}
The bounds of Eqs.\,(\ref{gagg_const}) and (\ref{gagg_constN}) are much lower than that of Eq.\,(\ref{k_bound}) coming from the present experimental upper bound ${\rm Br}(K^+\rightarrow\pi^+A_i)<7.3\times10^{-11}$\,\cite{Adler:2008zza} as well as the axion to electron coupling $6.7\times10^{-29}\lesssim\alpha_{Aee}\lesssim5.6\times10^{-27}$ at $3\sigma$\,\cite{Giannotti:2017hny}.

Hence, from Eqs.\,(\ref{k_bound}) and (\ref{axiMass2}) the QCD axion mass and its corresponding axion-photon couplings are model predicted for $z=0.48$:
  \begin{eqnarray}
  m_a=1.54^{+0.48}_{-0.29}\times10^{-4}\,\text{eV}\,\Leftrightarrow\,|g_{a\gamma\gamma}|=\left\{
       \begin{array}{lll}
         5.99^{+1.85}_{-1.14}\times10^{-14}\,\text{GeV}^{-1}, & \hbox{case-I} \\
         4.89^{+1.51}_{-0.93}\times10^{-14}\,\text{GeV}^{-1}, & \hbox{case-II}\\
         1.64^{+0.51}_{-0.31}\times10^{-14}\,\text{GeV}^{-1}, & \hbox{case-III}
       \end{array}
     \right.\,.
  \label{Amass_pre}
 \end{eqnarray}
Note here that, if $0.38<z<0.58$ is considered for the given axion mass range, the ranges of $|g_{a\gamma\gamma}|$ in Eq.\,(\ref{Amass_pre}) can become wider than those for $z=0.48$. 
The corresponding Compton wavelength of axion oscillation is $\lambda_a=(2{\pi\!\!\not\!h}/m_a)c$ with $c\simeq2.997\times10^{8}\,{\rm m/s}$ and ${\!\!\not\!h}\simeq1.055\times10^{-34}\,{\rm J}\cdot{\rm s}$:
 \begin{eqnarray}
  \lambda_a=8.04^{+1.90}_{-1.90}\,{\rm mm}\,.
 \end{eqnarray}
%%%%%%%%%%%%%%%
%    Fig 2    %
%%%%%%%%%%%%%%%
\begin{figure}[h]
%\begin{tabular}{c}
\includegraphics[width=11.0cm]{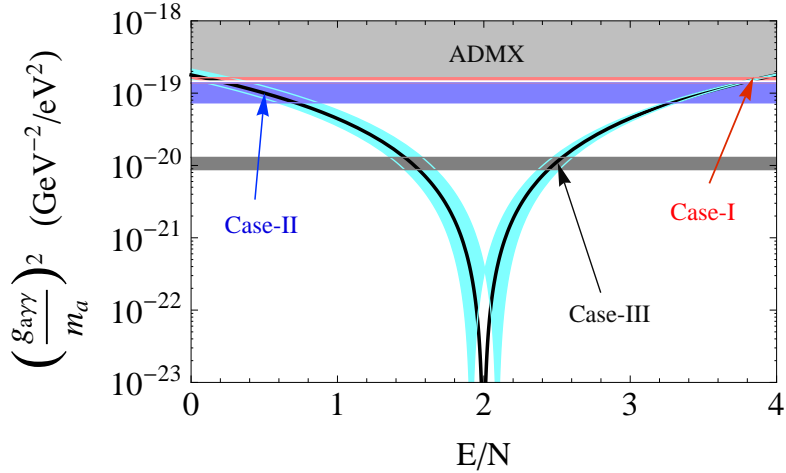}
%\end{tabular}
\caption{\label{Fig2} Plot of $(g_{a\gamma\gamma}/m_{a})^2$ versus $E/N$ for $z=0.48$ (black curve) and $0.38<z<0.58$ (cyon-band curve). The gray-band represents the experimentally excluded  bound $(g_{a\gamma\gamma}/m_{a})^2\leq1.44\times10^{-19}\,{\rm GeV}^{-2}\,{\rm eV}^{-2}$ from ADMX\,\cite{Asztalos:2003px, Asztalos:2009yp}. Here the horizontal light-red, light-blue, and light-black bands stand for $(g_{a\gamma\gamma}/m_{a})^2=1.507^{+0.126}_{-0.137}\times10^{-19}\,{\rm GeV}^{-2}\,{\rm eV}^{-2}$ for $E/N=23/6$, $1.003^{+0.382}_{-0.368}\times10^{-19}\,{\rm GeV}^{-2}\,{\rm eV}^{-2}$ for $E/N=1/2$, and $1.128^{+0.163}_{-0.252}\times10^{-20}\,{\rm GeV}^{-2}\,{\rm eV}^{-2}$ for $E/N=5/2$, respectively.}
\end{figure}
 The QCD axion coupling to photon $g_{a\gamma\gamma}$ divided by the QCD axion mass $m_{a}$ is dependent on $E/N$. Fig.\,\ref{Fig2} shows the $E/N$ dependence of $(g_{a\gamma\gamma}/m_{a})^2$ so that the experimental limit is independent of the axion mass $m_{a}$\,\cite{Ahn:2014gva}: for $0.38<z<0.58$, the value of $(g_{a\gamma\gamma}/m_{a})^2$ for the case-II and -III are located lower than that of the ADMX (Axion Dark Matter eXperiment) bound\,\cite{Asztalos:2003px}, while for the case-I is marginally\,\footnote{In fact, this is the case for $0.54\lesssim z<0.58$.} lower than that of the ADMX bound, where $(g_{a\gamma\gamma}/m_{a})^2_{\rm ADMX}\leq1.44\times10^{-19}\,{\rm GeV}^{-2}\,{\rm eV}^{-2}$. The gray-band represents the experimentally excluded bound $(g_{a\gamma\gamma}/m_{a})^2_{\rm ADMX}$, while the cyon-band curve stands for $0.38<z<0.58$.  For the Weinberg value $z=0.48^{+0.10}_{-0.10}$, the anomaly values $E/N=23/6$, $1/2$, and $5/2$ predict $(g_{a\gamma\gamma}/m_{a})^2=1.507^{+0.126}_{-0.137}\times10^{-19}\,{\rm GeV}^{-2}\,{\rm eV}^{-2}$ (case-I), $1.003^{+0.382}_{-0.368}\times10^{-19}\,{\rm GeV}^{-2}\,{\rm eV}^{-2}$ (case-II), and $1.128^{+0.163}_{-0.252}\times10^{-20}\,{\rm GeV}^{-2}\,{\rm eV}^{-2}$ (case-III), respectively.
Clearly, as shown in Fig.\,\ref{Fig2}, the uncertainties of $(g_{a\gamma\gamma}/m_{a})^2$ for the case-II and -III are larger than that of case-I for $0.38<z<0.58$.

%%%%%%%%%%%%%%%
%    Fig 3    %
%%%%%%%%%%%%%%%
\begin{figure}[h]
%\begin{tabular}{c}
\includegraphics[width=11.0cm]{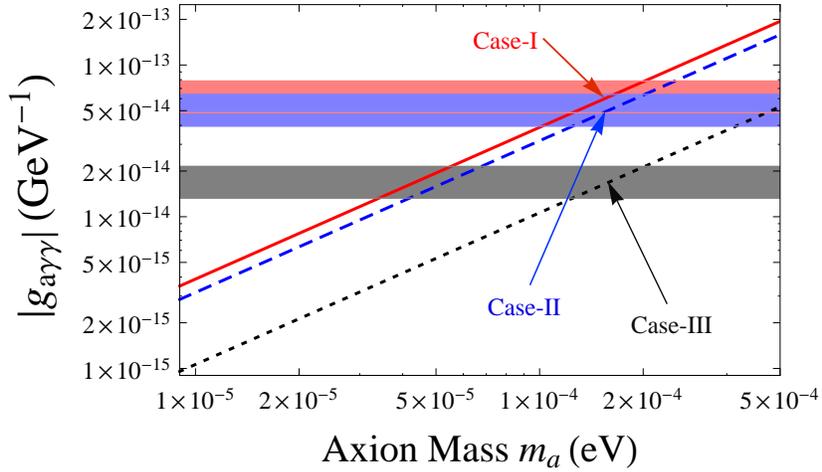}
%\end{tabular}
\caption{\label{Fig3} Plot of $|g_{a\gamma\gamma}|$ versus $m_{a}$ for the case-I (slanted red-solid line), case-II (slanted blue dashed line), and case-III (slanted black-dotted line) in terms of $E/N=$ $23/6$, $1/2$ and $5/2$, respectively. Especially, the QCD axion mass $m_{a}=1.54^{+0.48}_{-0.29}\times10^{-4}\,\text{eV}$ is equivalent to the axion photon coupling $|g_{a\gamma\gamma}|=5.99^{+1.85}_{-1.14}\times10^{-14}\,\text{GeV}^{-1}$ (horizontal light-red band), $4.89^{+1.51}_{-0.93}\times10^{-14}\,\text{GeV}^{-1}$ (horizontal light-blue band), and $1.64^{+0.51}_{-0.31}\times10^{-14}\,\text{GeV}^{-1}$ (horizontal light-black band), which corresponds to the case-I, -II, and -III, respectively.}
\end{figure}
Fig.\,\ref{Fig3} shows the plot for the axion-photon coupling $|g_{a\gamma\gamma}|$ as a function of the axion mass $m_{a}$ in terms of anomaly values $E/N=23/6, 1/2, 5/2$ which correspond to the case-I, -II, and -III, respectively. Especially, in the model, for $F_A=3.56^{+0.84}_{-0.84}\times10^{10}$ GeV and $z=0.48$ we obtain the QCD axion mass $m_{a}=1.54^{+0.48}_{-0.29}\times10^{-4}\,\text{eV}$ and the axion photon coupling $|g_{a\gamma\gamma}|=5.99^{+1.85}_{-1.14}\times10^{-14}\,\text{GeV}^{-1}$ (horizontal light-red band), $4.89^{+1.51}_{-0.93}\times10^{-14}\,\text{GeV}^{-1}$ (horizontal light-blue band), and $1.64^{+0.51}_{-0.31}\times10^{-14}\,\text{GeV}^{-1}$ (horizontal light-black band), which corresponds to the case-I, -II, and -III, respectively. 
As the upper bound on ${\rm Br}(K^+\rightarrow\pi^++A_i)$ gets tighter, the range of the QCD axion mass gets more and more narrow, and consequently the corresponding band width on $|g_{a\gamma\gamma}|$ in Fig.\,\ref{Fig3} is getting narrower. In Fig.\,\ref{Fig3} the top edge of the bands comes from the upper bound on ${\rm Br}(K^+\rightarrow\pi^++A_i)$, while the bottom of the bands is from the astrophysical constraints of star cooling induced by the flavored-axion bremsstrahlung off electrons $e+Ze\rightarrow Ze+e+A_i$.  

The model will be tested in the very near future through the experiment such as CAPP (Center for Axion and Precision Physics research)\,\cite{CAPP} as well as the NA62 experiment expected to reach the sensitivity of ${\rm Br}(K^+\rightarrow\pi^++A_i)<1.0\times10^{-12}$\,\cite{Fantechi:2014hqa}.

%%%%%%%%%%%%%%%%%%%%%%%%%%%%%%%%%%%%%%
\section{Summary and Conclusion}
Motivated by the flavored PQ symmetry for unifying the flavor physics and string theory\,\cite{Ahn:2016typ, Ahn:2016hbn}, we have constructed a compact model based on $SL_2(F_3)\times U(1)_X$ symmetry for resolving rather recent, but fast-growing issues in astro-particle physics, including quark and leptonic mixings and CP violations, high-energy neutrinos, QCD axion, and axion cooling of stars. Since astro-particle physics observations have increasingly placed tight constraints on parameters for flavored-axions,
we have showed how the scale responsible for PQ mechanism (congruent to that of seesaw mechanism) could be fixed, and in turn the scale responsible for FN mechanism through flavor physics. 
Along the lines of finding the fundamental scales, the flavored-PQ symmetry together with the non-Abelian finite symmetry is well flavor-structured in a unique way that domain-wall number $N_{\rm DW}=1$ with the $U(1)_X\times[gravity]^2$ anomaly-free condition demands additional Majorana fermion and the flavor puzzles of SM are well delineated by new expansion parameters expressed in terms of $U(1)_X$ charges and $U(1)_X$-$[SU(3)_C]^2$ anomaly coefficients, providing interesting physical implications on neutrino, QCD axion, and flavored-axion.

In the concrete, the QCD axion decay constant congruent to the seesaw scale, through its connection to the astro-particle constraints of stellar evolution induced by the flavored-axion bremsstrahlung off electrons $e+Ze\rightarrow Ze+e+A_i$ and the rare flavor-changing decay process induced by the flavored-axion $K^+\rightarrow\pi^++A_i$, is shown to be fixed at $F_A=3.56^{+0.84}_{-0.84}\times10^{10}$ GeV (consequently, the QCD axion mass $m_a=1.54^{+0.48}_{-0.29}\times10^{-4}$ eV, wavelength of its oscillation $\lambda_a=8.04^{+1.90}_{-1.90}\,{\rm mm}$, axion to neutron coupling $g_{Ann}=2.14^{+0.66}_{-0.41}\times10^{-12}$, and axion to photon coupling $|g_{a\gamma\gamma}|=5.99^{+1.85}_{-1.14}\times10^{-14}\,\text{GeV}^{-1}$ for $E/N=23/6$ (case-I), $4.89^{+1.51}_{-0.93}\times10^{-14}\,\text{GeV}^{-1}$ for $E/N=1/2$ (case-II), $1.64^{+0.51}_{-0.31}\times10^{-14}\,\text{GeV}^{-1}$ for $E/N=5/2$ (case-III), respectively, in the case $z=4.8$.). Subsequently, the scale associated to FN mechanism is automatically fixed through its connection to the SM fermion masses and mixings, $\Lambda=2.04^{\,+0.48}_{\,-0.48}\times10^{11}\,{\rm GeV}$, and such fundamental scale might give a hint where some string moduli are stabilized in type-IIB string vacua.

%We may conclude that, in an extended SM framework by a compact symmetry $G_F=SL_2(F_3)\times U(1)_X$ if the scale responsible for the FN mechanism (whose scale is associated to some string moduli stabilization) is fixed, the scales responsible for seesaw and PQ mechanisms can be {\it dynamically} determined in a way that the SM fermion (including neutrino) masses and mixings are well delineated, which in turn can provide predictions on several properties of the flavored-axions.
In the very near future, the NA62 experiment expected to reach the sensitivity of ${\rm Br}(K^+\rightarrow\pi^++A_i)<1.0\times10^{-12}$ will probe the flavored-axions or exclude the model.

%%%%%%%%%%%%%%%%%%%%%%%%%%%%%%%%%%%%%%%%%%%%%%%%%%%%%%%%%%%%%%%%%%%%%%%%%%%%%%%%%%%%%%%%%%%%%%%%%%%%%%%%%%%%%%%%
\acknowledgments{I would like to give thanks to BH An and MH Ahn for useful conversations. This work is supported under grant
no. XXXXX by the Institute of High Energy Physics,
Chinese Academy of Sciences.
}

%%%%%%%%%%%%%%%%%%%%%%%%%%%%%%%%%%%%%%%%%%%%%%%%%%%%%%%%%%%%%%%%%%%%%%%%%%%%%%%%%%%%%%%%%%%%%%%%%%%%%%%%%%%

\end{document}